\documentclass[aps,pre,twocolumn,superscriptaddress,showpacs]{revtex4-1}

  \usepackage{bm}
  \usepackage{graphicx}
  \usepackage{amsmath}

\begin{document}


\title{
  The limit of strong ion coupling due to electron shielding
  }

\author{M. Lyon}
\author{S. D. Bergeson}
\email{scott.bergeson@byu.edu}
\affiliation{Department of Physics and Astronomy, Brigham Young University, Provo, UT 84602, USA}
\author{M. S. Murillo}
\affiliation{Computational Physics and Methods Group, Los Alamos National Laboratory, Los Alamos, New Mexico 87545, USA}

\date{\today}

\begin{abstract}
  We show that strong coupling between ions in an
  ultracold neutral plasma is limited by
  electron screening. While electron screening reduces the
  quasi-equilibrium ion temperature,
  it also reduces the ion-ion electrical potential energy.
  The net result is that the ratio of nearest-neighbor potential
  energy to kinetic energy in quasi-equilibrium is constant and
  limited to approximately 2 unless the electrons are heated by some
  external source.
  We support these conclusions by reporting new measurements of the
  ion velocity distribution in an ultracold neutral calcium plasma.
  These results match previously reported simulations of
  Yukawa systems.
  Theoretical considerations are used to determine the screened
  nearest-neighbor potential energy in the plasma.
\end{abstract}

\pacs{52.27.Gr, 52.20.-j, 79.70.+q, 52.25.Jm}

\maketitle

\section{Introduction}

Ultracold neutral plasmas are strongly coupled Coulomb systems. Strong coupling occurs when the nearest-neighbor (Coulomb) potential energy is greater than the average kinetic energy per particle.
The degree of strong coupling between ions is traditionally characterized by the parameter
\begin{equation}
  \Gamma_{ii}^{} = \frac{Z^2 q^2}{4\pi\epsilon_0^{} a_{\rm ws}^{} k_{\rm B}^{} T}
  \label{eqn:gamma}
\end{equation}
where
$Z$ is the ion charge state, $q$ is the fundamental charge,
$a_{\rm ws}^{} = (3/4\pi n)^{1/3}$ is the Wigner-Seitz radius or the
average distance between ions in the plasma,
$n$ is the plasma density,
and $T$ is the temperature.
These plasmas are typically generated by photo-ionizing laser-cooled gases \cite{killian99}
or ions in an ultrasonic jet \cite{morrison08}.
They are diagnosed using three-body recombination \cite{killian01, fletcher07, bergeson08, pohl08, sadeghi11, bannasch11},
thermalization rates \cite{castro10, bergeson11, mcquillen11, lyon11},
electron evaporation or rf absorption \cite{killian01, bergeson03, roberts04, twedt10, lu11, twedt11, wilson12},
charged particle imaging and detection \cite{morrison08, zhang08},
and optical fluorescence \cite{cummings05a,lyon11,laha07}
and
absorption \cite{simien04, chen04, killian07}.
Theoretical calculations and simulations \cite{robicheaux02, kuzmin02a, pohl04a, murillo06,  park10, jin11, mendonca11, shukla11}
give great insights into the properties of these plasmas.

Achieving a higher value of the strong coupling parameter is a high priority in this field.
One of the primary objectives of ultracold plasma studies is understanding how strong coupling changes the description
of basic plasma or atomic processes. As the plasma becomes more strongly coupled, the fundamental
assumptions that make fluid approximations valid break down. Collisional processes that are well-understood
in terms of interactions between discreet atoms or ions in the plasma,
such as three-body recombination, should take on a many-body nature. While evidence for this
is seen for dense Rydberg gases \cite{park11}, only limited evidence for departures from traditional
plasma physics has been reported in the literature (see, for example \cite{bergeson08,bergeson11}).

At early times in the plasma evolution the ion temperature is dominated
by disorder-induced heating (DIH). Although ions are formed with essentially zero kinetic energy in ultracold neutral plasmas,
they have an excess of electrical potential energy due to the random spatial
distribution of nearest neighbors in the plasma.
In the absence of correlation and shielding effects in the final state, the initially $T\sim\mbox{mK}$ ions
heat up to the correlation temperature
\begin{equation}
  T_c^{} = \frac{2}{3} \frac{q^2}{4\pi\epsilon_0^{} a_{\rm ws}^{} k_{\rm B}^{}}.
\end{equation}
on the time scale of the inverse ion plasma frequency, $1/\omega_i^{} = (nq^2/m_i^{}\epsilon_0^{})^{-1/2}$, where $m_i^{}$ is the ion mass. In the absence of electron shielding,
the ion temperature is determined by the density alone.

In neutral plasmas, electrons form a screening background for the ions.
This changes the ion dynamics. If the
electron temperature is not too low, the ion interaction can be modeled with a Yukawa potential
  \begin{equation}
    u_{ii}^{Y}= \frac{q^2}{4\pi\epsilon_0}\frac{e^{\rm -r/\lambda_{\rm D}}}{r},
  \label{eqn:yukawa}
\end{equation}
where $\lambda_{\rm D} = (k_{\rm B}^{} T_{\rm e}^{} \epsilon_0^{} / nq^2)^{1/2}$ is the Debye length
and $k_{\rm B}^{}$ is Boltzmann's constant.
Electron screening can be parameterized using the inverse scaled screening length,
$\kappa \equiv a_{\rm ws}^{} / \lambda_{\rm D}^{}$. The parameter $\kappa$ exhibits a strong temperature
dependence, $\kappa \sim T_e^{-1/2}$, and a weak density dependence $\kappa \sim n^{1/6}$.

Previous studies have shown that electron screening can significantly reduce the ion temperature and slow the DIH time scale \cite{bergeson11, lyon11}. At first sight this appears to increase $\Gamma$. However, because electron
screening reduces the nearest-neighbor electrical potential energy, the net effect of electron shielding on the ratio of the actual nearest-neighbor potential energy to kinetic energy
is not immediately clear.

In this paper we present new fluorescence measurements of an ultracold neutral Ca plasma. From our measurements we extract the time evolving rms width of the velocity distribution by fitting the data to a Voigt profile. Using an expansion model we find the electron and ion temperature as a function of time. By varying the initial electron temperature we generate plasmas with varying degrees of electron shielding. We show that we can generate plasmas with very cold ions by mitigating the effects of DIH through electron shielding. We compare our experimental results to two molecular dynamics simulations, which show good agreement with each other and with our data.
We use theoretical considerations to extract the screened ion-ion potential energy in the plasma.
We suggest that although electron screening reduces heating due to DIH, it also reduces the nearest-neighbor potential energy in such a way that the ratio of potential energy to kinetic energy is independent of the electron temperature.

\section{Experiment}

We cool and trap approximately $20$ million $^{40}$Ca atoms in a
magneto-optical trap (MOT) at a temperature of a few mK.
The spatial density profile is Gaussian and has
the form $n(r) = n_0^{} \exp(-r^2/2\sigma^2)$, with peak density $n_0^{} \leq 2\times 10^{10}$ cm$^{-3}$ and
$\sigma = 0.2 - 0.4$ mm. We ionize these laser-cooled atoms using ns-duration laser pulses
at wavelengths of 423 and 390 nm that drive the $4s^2 \; ^1S_0 \rightarrow 4s4p \; ^1P_1^{\rm o}$ and the $4s4p \; ^1P_1^{\rm o}\rightarrow\mbox{continuum}$ transitions, respectively.

The initial ion and electron temperatures, $T_i^{}(0)$ and $T_e^{}(0)$, are controlled in the experiment but change in time. The initial ion temperature in the plasma is approximately
equal to the few mK temperature of the neutral atoms in the MOT. It changes as ions undergo DIH and as the plasma expands. The plasma is not confined by the MOT, and it freely
expands into the surrounding vacuum. The expansion velocity is typically
determined by electron temperature and the
ion mass. As long as the electron temperature is not too low, the
asymptotic expansion velocity is
$v_{\rm exp}^{} = (k_{\rm B}^{}T_{\rm e}^{} / m_i^{})^{1/2}$ \cite{gupta07}. The initial electron energy in the plasma is equal to the difference between the
ionizing laser photon energy and the calcium ionization energy. In our experiment this typically
ranges from $T_e^{} = 2E_e^{}/3k_B^{} = 0.5 - 150$ K.

We probe the ion velocity distribution using laser-induced fluorescence. We use a low-power cw laser
beam detuned from resonance at a wavelength of 397 nm, corresponding to the Ca$^+$ $4s~^2S_{1/2}^{}
\rightarrow 4p~^2P_{1/2}^{\circ}$ transition. The laser beam is
collimated to a diameter of 4~mm, attenuated to 2.5~mW, aligned to spatially overlap
the calcium plasma, and retroreflected. The maximum probe laser beam
intensity in the retroreflected configuration is approximately equal to the saturation
intensity, where the saturation intensity is $I_{sat} = 46$~mW/cm$^2$. Fluorescence photons at this same wavelength are collected
as a function of time after the plasma is generated using a 1-GHz bandwidth photo-multiplier tube and digital oscilloscope. We repeat this process
for a range of probe laser beam offset frequencies from 0 to about $\pm 250$~MHz.

\section{Data analysis}
\label{sec:analysis}

We extract the time evolving ion velocity $v_{i,\rm rms}^{}$ by fitting the fluorescence data to a Voigt profile. The Voigt profile is a mathematical representation of the absorption cross section per atom.  It is the convolution of a Lorentzian and a Gaussian lineshape
\begin{equation}
V(\nu) = \int_{-\infty }^{\infty}L(\nu-\nu')G(\nu')d\nu'
\label{eqn:voigt1}
\end{equation}
with the Lorenztian and Gaussian profiles given by
\begin{equation}
L(\nu) = \frac{\gamma/\pi}{\nu^2+\gamma^2}
\label{eqn:lorentzian}
\end{equation}
and
\begin{equation}
G(\nu) = \frac{1}{\sqrt{2\pi}\nu_{\rm rms}^{}}\exp[-\nu^2/2\nu_{\rm rms}^2],
\label{eqn:gaussian}
\end{equation}
respectively. In these equations, $\nu$ is the detuning from resonance, $\gamma$ is half the natural linewidth of the atomic transition (the HWHM of the Lorentzian line shape), and $\nu_{\rm rms}^{}$ is the rms Gaussian width. The integral defined in Eq. (\ref{eqn:voigt1}) can be evaluated as
\begin{equation}
V(\nu) = \frac{\textrm{Re}[w(z)]}{\sqrt{2\pi}\nu_{rms}}.
\label{eqn:voigt2}
\end{equation}
The term in the numerator is the complex error function, and it is given by
$w(z) = e^{-z^2} \mbox{erfc}\left( -iz \right )$, where z is $(\nu~+~i\gamma)/\sqrt{2}\nu_{\rm rms}^{}$, and erfc is the complementary error function. In the analysis, the Lorenztian half width is equal to 11~MHz, half the natural linewidth of the 397 nm transition. However it is power-broadened to 13.2 MHz (HWHM). The Gaussian width $\nu_{\rm rms}^{}$
is extracted as a fit parameter. It is converted to the rms width of the
velocity distribution using the Doppler shift,
$v_{\rm i,rms}^{} = (k_{\rm B}^{} T_i^{} /m_i^{})^{1/2} = \lambda \nu_{\rm rms}^{}$,
allowing us to map out the width of the ion velocity distribution as a function of time.

As the plasma evolves, the plasma ions are accelerated radially outwards.
The acceleration is well approximated by the expression
$a = (k_{\rm B}^{} T_e / m_i \sigma^2) r$ \cite{robicheaux02}.
This directed expansion adds to the thermal motion of the ions. In our experiment
we measure the total rms width of the velocity distribution, which contains
contributions from both the random thermal motion as well as the accelerated
expansion. Fortunately, the time scale for expansion is slower than the time
scale for DIH, and the contributions of each of these effects can be cleanly separated.

\begin{figure}
\centerline{\includegraphics[width=0.5\textwidth]{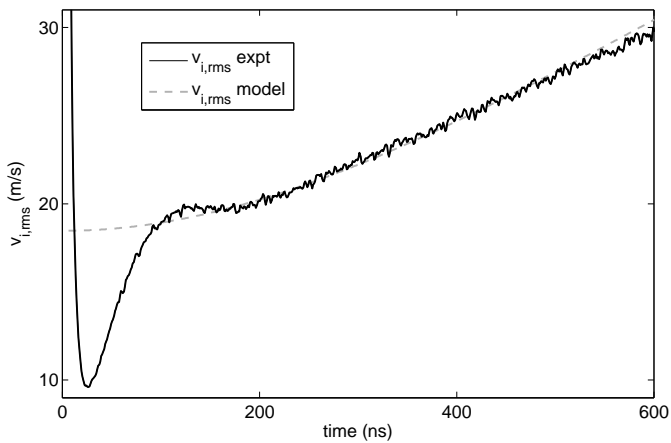}}
\caption{\label{fig:2}
  The time evolving rms width of the ion velocity distribution for an ultracold calcium plasma at a density of approximately $0.9\times 10^{10}$~cm$^{-3}$ and electron temperature of $T_e = 50~K$.  The rms velocity width is found using a fit to a Voigt profile, where the Gaussian frequency width is extracted as a fit parameter and converted to the velocity width through the Doppler shift. The model described by Eq. (\ref{eqn:laha}) is plotted as the gray dashed line.}
\end{figure}

Using the $v_{i,\rm rms}^{}$ found from the Voigt fitting we are able extract the ion temperature. The ion temperature can be related to the rms ion velocity as \cite{laha07}
\begin{equation}
v_{i,\rm rms}^{} = \sqrt{\frac{k_B^{}}{m_i^{}}\left\{\frac{t^2}{\tau_{\rm exp}^2}[T_e^{}(t) + T_i^{}(t)] + T_i^{}(t)\right\}},
\label{eqn:laha}
\end{equation}
where $\tau_{\rm exp}$, the characteristic expansion time, is given by $\tau_{\rm exp}~=~\sqrt{m_i \sigma(0)^2/k_{\rm B}[T_{\rm e}(0) + T_{\rm i}(0)]}$ and the time evolving ion and electron temperatures are given by $T_{\alpha}^{}(t) = T_{\alpha}^{}(0)/(1 + t^2/\tau_{\rm exp}^2)$ and the subscript $\alpha = i,e$. In Fig. \ref{fig:2} we have plotted the $v_{i,\rm rms}^{}$ calculated from this model and the $v_{i,\rm rms}^{}$ found by fitting a Voigt profile to the fluorescence data of one of our plasmas. Rearranging Eq. (\ref{eqn:laha}) we solve for the ion temperature $T_{i}(t)$
\begin{equation}
T_{i}(t) = \frac{m_i v_{\rm i,rms}^2}{k_{B}} - \frac{T_{e}^{}(0)\frac{t^2}{\tau_{\rm exp}^2}}{1 + \frac{t^2}{\tau_{\rm exp}^2}}
\label{eqn:laha2}
\end{equation}
 The initial ion temperature is assumed to be negligible compared to the initial electron temperature. On the 200~-~1000~ns timescale the plasma has not expanded.

 When the electron temperature is not known, it is possible to estimate its value using the expansion model in Eq.(\ref{eqn:laha}). This is useful for plasmas that evolve from cold Rydberg gases, for example, where the initial electron temperature is not well defined. At these small initial electron temperatures,
 the electron temperature extracted from the expansion model have been shown to overestimates the electron temperature at late times \cite{gupta07}. However, it is likely that the model is reasonably accurate at early times before the plasma has expanded.

\section{Experimental results}

\begin{figure}
\centerline{\includegraphics[width=0.5\textwidth]{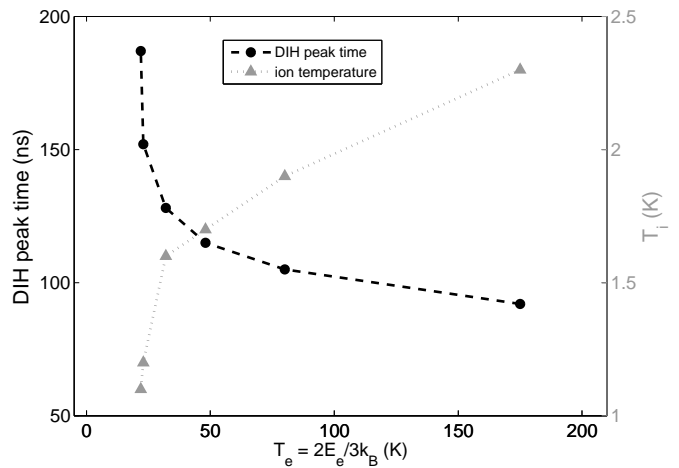}}
\caption{\label{fig:1a}
 The equilibrium ion temperature after DIH (triangles) and the characteristic DIH time (circles) plotted as a function of the electron temperature extracted from the expansion model. It is evident from this plot that electron screening reduces the ion temperature and extends the DIH time. The density for all these plasmas is approximately $0.9\times 10^{10}$~cm$^{-3}$}
\end{figure}

As the initial electron temperature decreases, the plasma expansion rate and ion thermalization rate changes. In Fig. \ref{fig:1a} we plot the average $T_i^{} = m_i^{} v_{\rm rms}^2/k_{\rm B}$ after the DIH process has completed. We also plot the characteristic DIH time for a range of electron temperatures. This characteristic time is taken to be the time when the ion temperature oscillation due to DIH is at its maximum. This can be seen in Fig. \ref{fig:2} at approximately 110 ns, where the measured $v_{i, \rm rms}^{}$ (black line) passes slightly above the model (gray dashed line). The electron and ion temperatures are both extracted from the expansion model given by Eq. (\ref{eqn:laha}). Plasmas with smaller $T_e^{}$ have smaller values of $T_i^{}$. The electron shielding length deceases with decreasing $T_e^{}$, softening the ion-ion Coulomb interaction. As the ions move under the influence of the screened Coulomb force of the neighboring ions, they acquire less kinetic energy compared to the unscreened case.

As $T_e^{}$ decreases, the time scale for DIH increases. This confirms observations in Refs. \onlinecite{bergeson11,lyon11}. The low-temperature electrons more effectively shield ions from their nearest neighbors. The Coulomb force is reduced, and the ions take longer to reach their ``equilibrium'' positions. The data in Fig. \ref{fig:1a} show that the DIH time is extended by as much as a factor of 2.

\section{Comparison with models}

Molecular dynamics simulations of complex neutral plasmas were published in the 1990's by Farouki, Hamaguchi, and Dubin \cite{farouki94, hamaguchi94, hamaguchi96, hamaguchi97}. Those simulations showed that electron shielding and correlation effects reduce the average electrical
potential energy of the plasma ions. Murillo showed that these simulations
can be applied to ultracold neutral plasmas \cite{murillo01}. A discussion of how that is done is given below.

\subsection{Deriving the DIH ion temperature}

The energy density per particle of the ultracold plasma can be generically written as
  \begin{equation}
    {\cal{E}} = \frac{3}{2}n k_B^{} \left(T_e^{} + T_i^{} \right) + \frac{n^2}{2} \sum_{a,b} \int d^3r \, u_{ab}(r) \, g_{ab}(r),
    \label{eqn:edens0}
  \end{equation}
where the summation indexes represent electrons $e$ or ions $i$. The Coulomb potential is written as $u_{ab}^{} = (q_a^{} q_b^{} / 4\pi\epsilon_0^{})(1/r)$ and the radial distribution function between species $a$ and $b$ is $g_{ab}^{}(r)$. This expression assumes a uniform plasma density, $n$. While the potential energy terms in Eq. (\ref{eqn:edens0}) are general, the kinetic energy is written in terms of the temperatures $T_e^{}$ and $T_i^{}$.

In order to quantify the ion heating during the DIH process, we will examine Eq. (\ref{eqn:edens0}) at two important instances in the plasma evolution. The first instance is just after the plasma is formed, after the electrons have thermalized with each other but before the ion have moved ($1/\omega_e^{} \sim 1\; \mbox{ns}$). Compared to the other energy scales in the system at this moment, the ions have essentially zero kinetic energy and we will therefore set the $T_i^{}$ to zero. We will call this instance the initial time.

The second instance is after the ions have thermalized with each other but before the plasma has expanded ($1/\omega_i^{} \sim 100-500 \; \mbox{ns}$). The ions will have moved primarily due to the Coulomb force of their nearest neighbors. Collisional transfer of energy from the electrons to the ions is much slower than the ion-ion collision rate. Consequently the electrons and ions maintain separate temperatures. The kinetic energy gained by the ions to this point will have come from the electrical potential energy of their screened neighboring ions. We will call this instance the final time.

Conserving energy, ${\cal{E}}_I^{} = {\cal{E}}_F^{}$, allows us to write,
  \begin{align}
    \left[ \frac{3}{2}n T_e^{} \right. & +
      \left.
      \frac{n^2}{2}
      \int d^3r \,
        \left(
          u_{ii}^{} g_{ii}^{} +
          u_{ee}^{} g_{ee}^{} +
          2u_{ei}^{} g_{ei}^{}
        \right)
    \right]_I^{} \nonumber \\
    &
    =
    \left[ \frac{3}{2}n \left(T_e^{}+T_i^{}\right) \right]_F^{} \nonumber \\
    &
    + \left[
      \frac{n^2}{2}
      \int d^3r \,
        \left(
          u_{ii}^{} g_{ii}^{} +
          u_{ee}^{} g_{ee}^{} +
          2u_{ei}^{} g_{ei}^{}
        \right)
    \right]_F^{},
    \label{eqn:edens00}
  \end{align}
where ${\cal{E}}_I^{}$ and ${\cal{E}}_F^{}$ are the initial and final energy densities and the explicit $r$-dependence of the potentials and distribution functions has been suppressed.

We can solve this equation if we assume that the electron temperature remains constant. In this case we can also ignore the term $u_{ee}^{}g_{ee}^{}$ on both sides of the equation because it remains unchanged from the initial to final time. We can also ignore the term $u_{ei}^{}g_{ei}^{}$ because the constant $T_e^{}$ approximation doesn't change the coupling between the electrons and ions when the ions move. The dominant change occurs in the ion-ion interaction.

With these approximations, Eq. (\ref{eqn:edens00}) becomes
  \begin{equation}
    \left[ \frac{3}{2}nT_i^{} \right]_F^{} =
    \left[
      \frac{n^2}{2}
      \int d^3r \, u_{ii}^{}
    \right]_I^{} -
    \left[
      \frac{n^2}{2}
      \int d^3r \, u_{ii}^{}g_{ii}^{}
    \right]_F^{},
    \label{eqn:edens000}
  \end{equation}
Note that the term $g_{ii}^{}$ in the initial state has been dropped because the initial state is completely disordered and $\left[ g_{ii}^{} \right]_I^{} = 1$. We can make a connection with the Yukawa-MD simulations of Refs. \onlinecite{farouki94, hamaguchi94, hamaguchi96, hamaguchi97} by introducing the Yukawa potential.
The final state ion-ion potential can be trivially written as
  \begin{equation}
    u_{ii}^{} = u_{ii}^Y + \left[ u_{ii} - u_{ii}^Y \right].
    \label{eqn:trivial}
  \end{equation}
Similarly, we can express the radial distribution function as
  \begin{equation}
    g_{ii}^{}(r) = \left[ g_{ii}^{}(r) -1 \right] + 1 = h_{ii}^{}(r) + 1,
  \end{equation}
where $h_{ii}^{}(r)$ is the pair correlation function.
Inserting these definitions into Eq. (\ref{eqn:edens000}) and simplifying gives
  \begin{align}
    \frac{3}{2}nT_i^{} = &
    - \frac{n^2}{2} \int d^3r \, u_{ii}^Y (h_{ii}+1) \nonumber \\
    &
    - \frac{n^2}{2} \int d^3r \, \left( u_{ii}^{} - u_{ii}^Y \right) h_{ii}^{}
    + \frac{n^2}{2} \int d^3r \, u_{ii}^Y
  \end{align}
where all quantities are evaluated in the final state. The second term on the RHS is small and can be neglected.  At small $r$, the quantity $u_{ii} - u_{ii}^Y$ is small and at large $r$ the pair correlation function $h_{ii}^{}(r)$ tends to zero. The last term on the RHS can be evaluated directly and is equal to $-(3/2)n\Gamma k_B^{} T_i^{}/\kappa^2$. The first term on the RHS has been tabulated using molecular dynamics (MD) simulations \cite{farouki94, hamaguchi94, hamaguchi96, hamaguchi97}. It is the potential energy of the Yukawa ions after the DIH process has completed.
In order to compare directly with the MD simulations, we need to convert from energy density to energy per particle. This is done by multiplying by the volume and dividing by the number of ions. We find the final ion temperature to be
  \begin{align}
    \frac{3}{2}
    k_B^{} T_i^{} = - \frac{n}{2} \int d^3r \, u_{ii}^Y (h_{ii}+1) - \frac{3\Gamma }{2\kappa^2} k_B^{} T_i^{}.
    \label{eqn:mdcomp}
  \end{align}
While this expression could be simplified further, we will leave it in this form in order to more easily compare with the results of previously published MD simulations.

The MD simulations by Hamaguchi tabulate the temperature-scaled
``excess energy'' per particle, $u \equiv \hat{U}/NkT$, which is written as \cite{hamaguchi96}
  \begin{align}
  u = \Gamma &
    \left[
      \frac{1}{N} \sum_{j=1}^{N-1} \sum_{k=j+1}^{N} \hat{\Phi}(|\vec{\xi}_k^{} - \vec{\xi}_j^{}|)
      -\frac{3}{2\kappa^2}
      -\frac{\kappa}{2} \right. \nonumber \\
      & \left.
      +\frac{1}{2}\sum_{n\neq 0}
        \frac{\exp\left(-\kappa \left|n\right|\Lambda\right)}{\left|n\right|\Lambda}
    \right]
    \label{eqn:hammy}
  \end{align}
We recognize the first term on the RHS of Eq. (\ref{eqn:hammy}) as the integral in Eq. (\ref{eqn:mdcomp}) divided by $k_B^{} T_i^{}$.
The last term on the RHS of Eq. (\ref{eqn:hammy}) explicitly accounts for the periodic boundary conditions, which we will neglect because we are considering an infinite-sized plasma. Equation (\ref{eqn:hammy}) includes the energy of the Debye sheath, $-\kappa/2$. To get the ``true'' potential energy per particle, we add this back in and multiply by $k_B^{} T_i^{}$,
  \begin{equation}
    \left(\frac{u}{\Gamma} + \frac{\kappa}{2} \right) k_B^{} T_i^{} \Gamma =
  \frac{\Gamma k_B^{} T_i^{}}{N} \sum_{j=1}^{N-1} \sum_{k=j+1}^{N} \hat{\Phi}(|\vec{\xi}_k^{} - \vec{\xi}_j^{}|)
      -\frac{3\Gamma}{2\kappa^2} k_B^{} T_i^{}
    \label{eqn:ham}
  \end{equation}
Comparing this with Eq. (\ref{eqn:mdcomp}) gives
  \begin{equation}
    T_i^{} = \frac{2}{3} \frac{q^2}{4\pi\epsilon_0^{} a_{ws} k_B^{}} \left( \frac{u}{\Gamma} + \frac{\kappa}{2} \right).
    \label{eqn:dihtemp}
  \end{equation}
This derivation is complementary to the one presented in \cite{murillo01}. It's presentation here provides information regarding the approximations and assumptions used in obtaining this result.

\subsection{Experimental and theoretical DIH ion temperature}

\begin{figure}
\centerline{\includegraphics[width=0.5\textwidth]{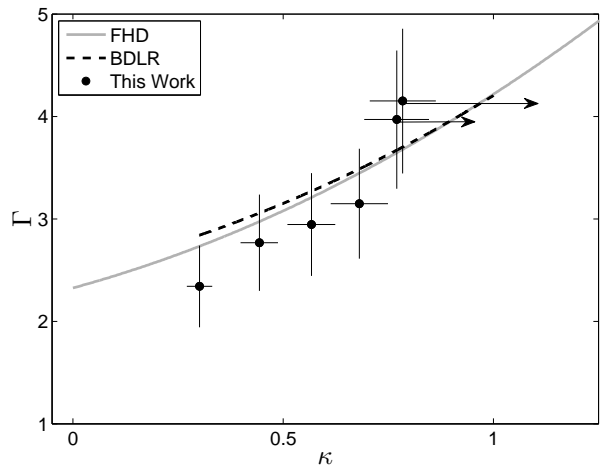}}
\caption{\label{fig:3}
Theoretical and experimental plots of the coupling parameter $\Gamma$ as a function of the electron screening $\kappa$. The solid gray line is derived from MD simulations \cite{farouki94, hamaguchi94, hamaguchi96, hamaguchi97}. The dotted black line is from a recent MD simulation of an ultracold plasma \cite{bergeson11}. The results of this work are also plotted as black circles with estimated error bars. The two rightmost experimental data points correspond to plasmas with low initial electron temperatures, as described in the text. Under these conditions, the model of Eq (\ref{eqn:laha2}) tends to overestimate the electron temperature, as suggested by the arrows.}
\end{figure}

Using Eq. (\ref{eqn:dihtemp}) and the MD results of Hamaguchi et al. \cite{farouki94, hamaguchi94, hamaguchi96, hamaguchi97}, we can predict the ion temperature after the DIH process has completed. This determination requires an iterative process because the ion temperature appears in the RHS of Eq. (\ref{eqn:dihtemp}) in the $u/\Gamma$ term \cite{chen04}. One begins by choosing an initial ion temperature, density, and electron $\kappa$. From this the ion $\Gamma$ can be calculated. The tables in Refs. \onlinecite{farouki94, hamaguchi94, hamaguchi96, hamaguchi97} then give the value of $u/\Gamma$. This can be inserted into Eq. (\ref{eqn:dihtemp}) which gives a new ion temperature. The process is repeated until the ion temperature and $\Gamma$ converge to a self-consistent limit. The resulting $\Gamma$ as a function of $\kappa$ is plotted in Fig. \ref{fig:3}. As $\kappa$ increases, the Debye length $\lambda_D^{}$ becomes smaller. The electrons more effectively shield the neighboring ions from one another and the final DIH temperature decreases. The $\Gamma$ vs. $\kappa$ plot is a plot of $1/T_i^{}$ vs. $1/\sqrt{T_e^{}}$.

A more recent MD simulation was published that calculated the influence of electron shielding on the ion DIH temperature \cite{bergeson11}. This MD simulation of Yukawa-shielded calcium ions started with the ions at rest, randomly positioned in a cell. The ions were allowed to move in the field generated by all of the other shielded ions in the cell (with periodic boundary conditions). After several $\omega_i^{-1}$, the average ion kinetic energy was calculated and from this the ion temperature was determined. The result is given in Eq. (6) in Ref. \onlinecite{bergeson11}. This is plotted as a dashed line in Fig. \ref{fig:3}. The $\kappa$ domain of this calculation is somewhat limited because the plasma becomes non-ideal when $\kappa > 1$ in the singly-ionized plasma used in the study, and in those conditions for that plasma it is not clear that the Yukawa approximation is valid. The excellent agreement between this result and the predictions based on Eq. (\ref{eqn:dihtemp}) is readily apparent.

In Fig. \ref{fig:3} we also plot our experimental results. The experimental determination of the electron and ion temperatures is described in Sec. \ref{sec:analysis}. There is excellent agreement between the experimental data and the two simulations described above. The rightmost experimental data point is measured in a plasma evolving from a Rydberg gas excited $\approx 10$~cm$^{-1}$ below the ionization potential. The second rightmost point corresponds to a plasma excited right at threshold. The expansion model we used to find the electron and ion temperatures for these two plasmas tends to overestimate the electron temperature at the very early times \cite{gupta07}. Thus we would expect the actual electron temperature to be lower, corresponding to larger values of $\kappa$, as suggested by the arrows in the plot.

The data in Fig. \ref{fig:3} demonstrate the validity of the assumptions used in deriving Eq. (\ref{eqn:dihtemp}). The data show that electron screening substantially reduces the ion temperature, resulting in increased values of $\Gamma_{ii}^{}$. Electron screening significantly mitigates the effects of DIH, which is the source of ion heating at these early times.  Our lowest temperature plasmas have $\Gamma_{ii}^{} = 4$.

\subsection{Screened potential energy vs. screened ion temperature}

In non-neutral plasmas, the parameter $\Gamma_{ii}^{}$ completely defines the ion-ion interactions. However, in \textit{neutral} plasmas, a ion-ion interaction necessarily includes contributions from the electrons. When the shielding length becomes comparable to the distance between ions, when $\kappa \rightarrow 1$, the relevance of $\Gamma_{ii}^{}$ is questionable.

One might be tempted to look at Eq. (\ref{eqn:yukawa}) and assume that the ``effective'' coupling constant is $\hat{\Gamma} = \Gamma_{ii}^{} \exp\left({-\kappa}\right)$. However, that would overestimate the influence of screening. For small $\kappa$, corresponding to the limit of weak screening, the first-order correction in that model would be linear in $\kappa$. This is clearly not the case, as MD simulations show. For example, the ion temperature and density at the liquid-solid phase transition clearly has no linear term [see Fig. 1 in Ref. \onlinecite{hamaguchi96}].

The idea of calculating $\Gamma_{ii}^{}$ is somewhat problematic in neutral plasmas. The $\Gamma$ parameter is supposed to represent the ion-ion nearest neighbor potential energy divided by the ion temperature. The problem arises because the ions and electrons are also correlated, and $\Gamma_{ei}^{}$ becomes important [see Eq. (\ref{eqn:edens0})]. There is a potential energy associated with $\Gamma_{ei}^{}$ that is shared by both the electrons and the ions. Similarly, because the electrons follow the ions, there is also a $\Gamma_{ee}^{}$ term that becomes important and that will mimic the $\Gamma_{ii}^{}$ behavior. When trying to calculate the screened ion coupling parameter, $\hat{\Gamma}$, it is not immediately clear which potential energy is appropriate to include in the calculation. They are all important and they all are connected to the ion density and temperature.

This distinction is important to make. The thermodynamic properties of non-neutral plasmas depend on $\Gamma_{ii}^{}$. These properties can be translated into the realm of neutral plasmas with the idea that weak electron screening modifies them only slightly. However in ultracold neutral plasmas where $\kappa =1$ is achievable, the $\Gamma_{ii}^{}$ scaling of these properties is not immediately clear. This is particularly the case when the $\Gamma_{ii}^{}$ is determined by $\kappa$, such as we show in Fig. \ref{fig:1a}.

In light of the fact that all of the electron and ion coupling parameters are important and interconnected, we can simply define $\hat{\Gamma}$ to be the total potential energy of the system divided by the kinetic energy of the ions. Taking $U$ as the total potential energy and $K$ to be kinetic energy of the ions, we write
  \begin{equation}
    \hat{\Gamma} = \frac{U_F^{}}{K_F^{}} = \frac{U_F^{}}{U_I^{} - U_F^{} + K_I^{}},
    \label{eqn:econs0}
  \end{equation}
where the conservation of energy is, trivially, $U_I^{}+K_I^{}=U_F^{}+K_F^{}$ and we have assumed that the electron temperature does not change from the initial to final state. Because the ions start out with mK temperatures, we can set $K_I^{}=0$. Summing up the contributions of the electrons and ions to the total initial potential energy gives $U_I^{} = 0$. This can be seen in two ways. One is that the initial state is completely uncorrelated and neutral and therefore the total potential energy must be zero. The other is to argue that the electron-ion potential energy terms are negative and exactly cancel the electron-electron and ion-ion potential energy terms. Either way, we end up at the conclusion that the magnitude of the screened coupling parameter is
  \begin{equation}
    \hat{\Gamma}=1.
    \label{eqn:ghat}
  \end{equation}
Even though electron screening reduces the ion temperature (see Fig. \ref{fig:1a}), it reduces the potential energy by exactly the same amount so that the ratio of potential energy to kinetic energy is always 1.

This result, Eq. (\ref{eqn:ghat}), is true for all neutral systems in which there is no external source of heat for the electrons and when there is no correlation in the initial state. The agreement between the experimental data and MD simulations in Fig. \ref{fig:1a} suggest that three-body recombination and electron-Rydberg scattering have not significantly increased the electron temperature at these early times, because those heating terms are not included in the MD simulation. If the electrons are heated, then the potential energy $U_F^{}$ in Eq. (\ref{eqn:econs0}) goes down and $\hat{\Gamma}$ will increase.

We note that the final state of the plasma cannot be completely determined by energy conservation alone because of the two-temperature nature of the UNP. For a given initial energy, there are many possible values of the final temperatures $T_e^{}$ and $T_i^{}$ that correspond to a correct final energy, at least in principle. Of course, if a true equilibrium state could be reached, the plasma would have $T_i^{} = T_e^{}$ and the final state would be deterministic. This suggests that more work on the quasithermodynamics of two-temperature plasmas is warranted \cite{dharma08,seuferling89}.

\section{Conclusion}

In this paper we present experimental measurements of laser-induced fluorescence from the ions in an ultracold neutral plasma. Fluorescence measurements are made when the probe laser frequency is scanned over the emission lineshape. From these fluorescence data we extract the rms velocity distribution as a function of time by fitting the data to a Voigt profile. An expansion model is used to find the electron and ion temperatures. This is done over a range of initial electron temperatures, which allows us to study the effect of electron shielding on ion equilibration at early times. Information about the ion and electron temperatures is used to calculate the strong coupling parameter $\Gamma_{ii}^{}$ and the electron shielding $\kappa$. We compare our experimental results with molecular dynamics simulations and theoretical calculations for the ion strong coupling in ultracold plasmas as a function of the electron shielding. We find that our experimental data show good agreement with MD results. We generate plasmas with very cold ions because electron screening mitigates the effects of DIH. However we also find that electron shielding softens the ion interaction strength, which has the net effect of keeping the ratio of potential energy to kinetic energy constant for all values of $\kappa$.

Our results indicate that it may be possible to use electron screening to generate a strongly coupled plasma
with $\Gamma_{ii}^{} >4$. This could be done by ionizing a low-density atom cloud with very low
initial electron temperature. The low density will reduce the time scale for electron
heating due to three-body recombination. The plasma electrons could be
heated so that the ions are adiabatically shifted into their equilibrium positions
in an unscreened plasma as $\kappa$ is reduced to zero. A large initial size for the plasma
would also reduce the time scale for the plasma to expand.

\section{Acknowledgements}

This work is supported in part by the National Science Foundation (grant no PHY-0969856) and the Air Force (grant no. FA9950-12-1-0308).

\end{document}